# Fluorographynes: Stability, Structural and Electronic Properties

A. N. Enyashin, A. L. Ivanovskii *

*Institute of Solid State Chemistry, Ural Branch of the Russian Academy of Sciences, Ekaterinburg, GSP-145, 620990, Russia*

A B S T R A C T

The presence in the graphyne sheets of a variable amount of $sp^2/sp^1$ atoms, which can be transformed into $sp^3$-like atoms by covalent binding with one or two fluorine atoms, respectively, allows one to assume the formation of fulorinated graphynes (fluorographynes) with variable F/C stoichiometry. Here, employing DFT band structure calculations, we examine a series of fluorographynes, and the trends in their stability, structural and electronic properties have been discussed as depending on their stoichiometry: from $C_2F_3$ (F/C= 1.5) to $C_4F_7$ (F/C= 1.75).

* Corresponding author: Institute of Solid State Chemistry, Ural Branch of the Russian Academy of Sciences, Ekaterinburg, Pervomaiskaya Str., 91, 620990 Ekaterinburg, Russia. Fax:+7(343)3744495

E-mail address: ivanovskii@ihim.uran.ru (A.L. Ivanovskii)

## 1. Introduction

Graphene, a two-dimensional (2D) mono-atomic-thick sheet of $sp^2$ hybridized carbon, exhibits a unique combination of structural, mechanical, electronic, thermal, and possibly magnetic properties, and is viewed today as a very promising material for various applications in a vast range of nanotechnologies, see reviews [1-5]. However "graphene is not the end of the road" [6]; and numerous efforts have been focused recently on the search of graphene-based materials with novel functionalities, in particular, through adsorption of various atoms or molecules on the surface of grapheme [7,8]. For example such atoms as fluorine, oxygen, or hydrogen adsorbed on graphene can form covalent bonds with the carbon atoms, which lead to the change of the hybridization state from $sp^2$ to $sp^3$ and may provoke the opening of a band gap.

In this way, fluorination of graphene can give rise to wide-band-gap 2D crystals, which were termed *fluorographenes* [9]. A set of outstanding chemical and physical properties for single-layer fluorographenes has been recently found experimentally and predicted theoretically, see Refs. [8-17] and References therein. These materials (with a variable F/C content – up to stoichiometry CF) are considered to provide a promising platform for interesting applications.

At the same time, the versatile flexibility of carbon to form various competing hybridization states allows one to design numerous types of flat single-atom-thick carbon networks: so-called graphene allotropes [18]. One of interesting families of these allotropes is represented by so-called *graphynes* (GPs), which can be described as graphene lattices, where some =C=C= bonds are replaced by uniformly distributed acetylenic linkages (-C≡C-) [19].

These carbon ($sp^2$+$sp^1$) sheets with a high level of π-conjunctions, with uniformly distributed pores, and with the density much less than that of graphene, possess unusual electronic properties, nonlinear optical susceptibility, thermal resistance, conductivity, and through-sheet transport of ions. They are considered as possible promising materials for nanoelectronics, for hydrogen storage, as membranes (for example, for hydrogen separation from syngas - as an alternative for nanomeshy graphene), for energy storage applications or as candidates for the anode material in batteries [20-26]. Thus, the tuning of the properties of these materials should be critical for their further applications.

Herein, we explore theoretically the structural, electronic properties and stability of series of hypothetical fluorinated graphynes (*fluorographynes,* FGPs) which can become likely candidates for the engineering of novel graphyne-based materials.

## 2. Computational aspects

All the calculations were performed by means of the density functional theory (DFT) [27] using the SIESTA 2.0 code [28,29] within the local-density approximation (LDA) with the exchange–correlation potential in the Perdew-Zunger form [30]. The core electrons were treated within the frozen core approximation using norm-conserving Troullier–Martins pseudopotentials [31]. The valence electrons were taken to be $2s^22p^2$ for C and $2s^23p^4$ for F. The pseudopotential core radii were chosen as suggested by Martins, and were equal to 1.50 and 1.54 bohr for *s*- and *p*-states of C, and 1.30 bohr for both *s*- and *p*-states of F. In all the calculations, only a single-$\zeta$ basis set was used for all atoms.

For *k*-point sampling, a cutoff of 10 Å was used [32]. The *k*-point mesh was generated by the method of Monkhorst and Pack [33]. A cutoff of 350 Ry for the real-space grid integration was utilized. All the calculations were performed using variable-cell and atomic position relaxation, with convergence criteria set to correspond to the maximum residual stress of 0.1 GPa for each component of the stress tensor, and the maximum residual force component of 0.01 eV/Å. Initial interlayer spacing along *c*-direction of a hexagonal or a rectangular lattice was set to 50 Å.

## 3. Structural models and results

At the first step we verified the utilized computational approach to probe of the fluorinated carbon systems on the example of well studied fluorographene. For this purpose, four isomers of fully fluorinated graphene (with stoichiometry CF) depicted in Fig. 1 were examined. The results obtained are listed in Table 1 and demonstrate that (i) all CF isomers are wide-band-gap semiconductors; (ii) the *chair* configuration is the most stable among all the examined isomers; and (iii) the stability of the isomers decreases in the sequence: *chair* (*1*) > *boat* (*2*) > *zigzag* (*3*) > *armchair* (*4*), whereas the band gap values in this sequence increase. These results (as well as the lattice parameters of CF isomers, see Table 1) are in good agreement with those obtained within other DFT-based calculations [14,16,17].

In forthcoming simulation of FGPs we considered three basic carbon graphyne sheets GP*1* – GP*3* depicted in Fig. 2 with quite different properties, namely:

(i). *Structural properties*: these networks comprise various possible combinations of $sp^1$- $sp^2$ atoms. E.g., the network GP*1* includes the hexagons $C_6$ formed by $sp^2$ atoms (C=C bonds) which are interconnected by -C≡C- linkages; the network GP*2* comprises the pairs of $sp^2$ atoms and finally in the network GP*3* only the isolated $sp^2$ atoms are presented.

(ii). *Electronic properties*: our preliminary calculations of GP*1* - GP*3* demonstrate (Figure 2, see also [18-20,22,26]) that these systems hold very different electronic properties: the network GP*1* (called also as 6,6,6-graphyne [19]) behaves as a semiconductor (with BG ~ 0.54 eV), whereas the band structures of the network GP*2* (called also as 14,14,14-graphyne [19]) and the network GP*3* (called also as 18,18,18-graphyne [19] or α-graphyne [26]) seem very intriguing: the valence and the conduction bands meet in a single point at the Fermi level, forming the Dirac cones. Accordingly the density of states (DOS) for GP*2,3* is zero at the Fermi level and these graphynes behave as a semiconductors with a zero band gap - like graphene [2,6].

(iii) *Stability*: According to our estimations (Table 2), the stability of the networks GP*1* - GP*3* strongly depends on the ratio of numbers of $sp^2/sp^1$ atoms. The greater the number of two-fold coordinated $sp^1$ atoms, the lower stability is in the sequence: GP*1* > GP*2* > GP*3*.

On the example of these GP*1* - GP*3* the main changes in their stability, structural and electronic properties by fluorination are examined. This choice allows to simulate the properties of fluorographynes depending on their stoichiometry. Indeed, herein we discuss the "fully fluorinated" graphynes, where all $sp^2$ and $sp^1$ atoms are transformed into the $sp^3$ hybridization state owing to the binding with one or two fluorine atoms, respectively. Full saturation of double- and triple C-C bonds of the networks GP*1,* GP*2,* and GP*3* bears the fluorographynes with stoichiometry $C_2F_3$ (F/C= 1.5), $C_3F_5$ (F/C= 1.66), and $C_4F_7$ (F/C= 1.75), respectively. Let us note that the degree of fluorination for graphyne-like sheets (i) can be higher than for graphene (the maximal ratio F/C=1), and (ii) differ depending on the ratio of $sp^2/sp^1$ atoms in different GP networks.

For every type of fluorographyne a series of possible ordered isomers and conformers (seven for FGP-$C_2F_3$, eight for FGP-$C_3F_5$, and six for FGP-$C_4F_7$) with different types of arrangement of F atoms has been constructed. Clearly, a wealth of FGP structures exists at every composition. Already within single unit cell of a fluorographyne a set of isomers may be constructed by the variation of the ratio of F atoms chemisorbed on $sp^2$ C atoms from different sides of source carbon sheet (≡C-F fragments). Fluorination of bridge $sp^1$ C atoms in a graphyne provides the existence of numerous conformers due to possible rotation and different relative orientation of difluoromethylene units within occuring –$CF_2$–$CF_2$– bridges. However, taking into account the classical concepts of steric effects, the number of initial geometries of fluorographynes within every type can be limited. We consider only the sets of variants near the maximal number of staggered *anti*-conformations of –$CF_2$–$CF_2$– bridges and with F atoms evenly spaced from both sides of carbon sheet. The optimized structures for all of 21 studied isomers of FGPs are available in *Supporting Information*.

In Table 3 the optimized lattice constants and formation energies ($E_{form}$) of the studied fluorographynes are listed. Here, the theoretical $E_{form}$ were estimated assuming the formal reactions: $nC(graphene) + \frac{1}{2}F_2 = C_nF$, and the values of $E_{form}$ were calculated as: $E_{form}(C_nF) = [E_{tot}(C_nF) - \frac{1}{2}E_{tot}(F_2) - nE_{tot}(C(graphene))]$, where $E_{tot}$ are the total energies of the corresponding substances as obtained in our calculations. Within this definition a negative $E_{form}$ indicates that it is energetically favorable for given reagents to form stable phases, and *vice versa.*

We see that except two isomers of FGP-$C_2F_3$, all other types of fluorographynes adopt negative values of $E_{form}$. Besides, (i) the differences in $E_{form}$ between various isomers for each fluorographyne as going from FGP-$C_2F_3$ to FGP-$C_4F_7$ (*i.e.* with the growth of the ratio F/C) decrease, and (ii) for the most stable isomers of each fluorographyne, their theoretical $E_{form}$ increase in the sequence: FGP-$C_2F_3$ < FGP-$C_3F_5$ < FGP-$C_4F_7$. Thus we can conclude that with the growth of the ratio F/C the stability of the fluorinated graphyne networks increases.

A comparison of relative formation energies for the isomers and conformers of fluorographynes of the same stoichiometry corroborates, that the steric effects should play a major role in the stability of these molecular networks. Within every class of FGPs the first isomer can be distinguished as the most stable isomer (Table 3 and *Supporting Information*). These isomers should possess minimal strain energy of the layers due to the ratio of F atoms chemisorbed from different sides as equal to 0.5. Moreover, the ordering of F atoms within these structures can be characterized as strongly alternant at all ≡C-F fragments and as always *anti*-conformic at –$CF_2$-$CF_2$- bridges, which provides the minimal energy of the steric stress.

Let us discuss the electronic properties of the examined FGPs on the example of their most stable isomers. From the electronic band pictures (Fig. 3) we see that: (i) all of the fluorographynes, irrespective of the type of the electronic spectrum of the parent

GP networks, behave as wide-band-gap semiconductors with direct $\Gamma$-$\Gamma$ inter-band transitions, (ii) the calculated LDA gaps increase in the sequence: FGP-$C_2F_3$ (3.36 eV) < FGP-$C_3F_5$ (3.64 eV) < FGP-$C_4F_7$ (4.27 eV), and (iii) the gaps for the examined fluorographynes become larger than for the fluorographenes as estimated within the same computational approach, see Table 1.

## 4. Conclusions

In summary, we have investigated the trends in stability, structural, and electronic properties of the proposed fluorinated graphynes (fluorographynes) with variable F/C stoichiometry (in examined case from $C_2F_3$ (F/C= 1.5) up to $C_4F_7$ (F/C= 1.75)), *i.e.* much higher than F/C ratio for graphene (F/C=1), which may be likely candidates for engineering of novel 2D materials.

Our DFT calculations have revealed important effects of fluorination of graphyne sheets. The higher F/C ratio, the stability of the fluorographynes increases, whereas the stability differences between their possible isomers decreases. Let us also note that the increase of the amount of $sp^1$ atoms in graphyne sheets influences on the stability of "pure" graphynes and their fuorinated derivatives in opposite ways, providing the reduction or the growth of their stability, respectively. Besides, we found that irrespective of the type of the electronic spectrum of "pure" graphynes, their fluorinated derivatives became wide-band-gap semiconductors with direct $\Gamma$-$\Gamma$ inter-band transitions, and the gaps increased with growth of the ratio F/C. Thus, the degree of fluorination, which is closely related with the actual ratio of $sp^1/sp^2$ atoms in graphyne sheets, is the main factor that stabilizes the fluorographynes and to regulates the dielectric properties of these materials.

We hope that these results will open interesting prospects for further experimental and theoretical studies of the related new materials. This especially concerns the fluorinated forms of very recently synthesized [34] grapdiynes with diacetylenic linkages - *i.e.* with an enhanced ratio of $sp^1/sp^2$ atoms as compared with the examined graphynes. This should promote an even greater degree of fluorination of these carbon sheets.

**Table 1.**
Calculated lattice parameters (*a* and *b*, in Å), relative energies (ΔE$_r$, in eV per carbon atom) and band gaps (BG, in eV) for four isomers of fluorographene CF in comparison with available data.

| isomer | *a* | *b* | ΔE$_r$ | BG | References ** |
|---|---|---|---|---|---|
| *chair* (*1*) * | 2.592 | 4.488 | 0 | 2.77 | |
| *boat* (*2*) | 2.587 | 4.530 | 0.093 | 2.82 | Present work |
| *zigzag* (*3*) | 2.606 | 4.117 | 0.082 | 3.26 | |
| *armchair* (*4*) | 4.848 | 4.539 | 0.130 | 3.29 | |
| *chair* | 2.600 | 4.503 | 0 | 3.20 / 7.42 *** | |
| *boat* | 2.574 | 4.602 | 0.075 | 3.23 / 7.32 | Ref. [14] |
| *zigzag* | 2.625 | 4.183 | 0.036 | 3.59 / 7.28 | |
| *armchair* | 4.886 | 4.231 | 0.095 | 4.23 / 7.98 | |
| *chair* | 2.61 | 4.520 | 0 | 3.10 | |
| *boat* | 2.58 | 4.570 | 0.150 | 3.28 | Ref. [16] |
| *stirrup* | 2.63 | 4.220 | 0.070 | 3.58 | |
| *twist-boat* | 5.14 | 4.600 | 0.150 | 3.05 | |
| *chair* | 2.611 | 4.521 | 0 | - | |
| *bed* | 2.585 | 4.617 | 0.148 | - | Ref. [17] |
| *washboard* | 2.635 | 4.200 | 0.071 | - | |

* Atomic configurations of isomers *1-4* see Figure 1;
** Other theoretical calculations: Ref. [14] -, Ref. [16]: DFT-LDA/GW, and Ref. [17]: DFT-PBE (ABINIT code);
*** as obtained within LDA/GW approximations

**Table 2.**
The calculated lattice constants (*a* and *b*, in Å), band gaps (BG, in eV) and relative energies (ΔE$_r$, in eV per atom) for graphynes (*1-3*).

| system * | lattice type | lattice constants | BG | ΔE$_r$ ** |
|---|---|---|---|---|
| *1* | hexagonal | *a* = 12.23 | 0.54 | 0.938 |
| *2* | rectangular | *a* = 7.22; *b* = 7.01 | 0 | 1.241 |
| *3* | hexagonal | *a* = 7.20 | 0 | 1.416 |

* see Fig. 1;
** relative to E$_{tot}$ of graphene

**Table 3.**
Calculated lattice parameters (*a* and *b*, in Å) and formation energies ($E_{form}$, in eV) for the examined isomers of fluorographynes.

| fluorographine | isomer * | *a* | *b* | $E_{form}$ |
|---|---|---|---|---|
| FGP-$C_2F_3$ | **1** | **13.147** | **13.143** | **-0.6831** |
|  | 2 |  | *decomposed* |  |
|  | 3 | 12.900 | 12.900 | -0.5872 |
|  | 4 |  | *decomposed* |  |
|  | 5 | 13.440 | 13.439 | -0.2334 |
|  | 6 | 13.072 | 12.742 | -0.4413 |
|  | 7 | 13.415 | 13.242 | -0.3816 |
| FGP-$C_3F_5$ | **1** | **6.943** | **7.905** | **-0.9155** |
|  | 2 | 8.241 | 5.901 | -0.6753 |
|  | 3 | 6.493 | 7.340 | -0.7938 |
|  | 4 | 7.022 | 7.975 | -0.8121 |
|  | 5 | 8.499 | 6.783 | -0.4864 |
|  | 6 | 6.939 | 7.612 | -0.8062 |
|  | 7 | 7.066 | 8.108 | -0.6939 |
|  | 8 | 7.049 | 7.690 | -0.8269 |
| FGP-$C_4F_7$ | **1** | **6.667** | **6.685** | **-1.0531** |
|  | 2 | 7.141 | 6.561 | -1.0433 |
|  | 3 | 7.294 | 6.333 | -0.9800 |
|  | 4 | 5.722 | 6.607 | -0.9996 |
|  | 5 | 5.089 | 7.109 | -1.0360 |
|  | 6 | 7.271 | 7.160 | -0.9220 |

* The structures for all of 21 studied isomers are given in *Supplementary data* as Figs S1-S3.

# FIGURES

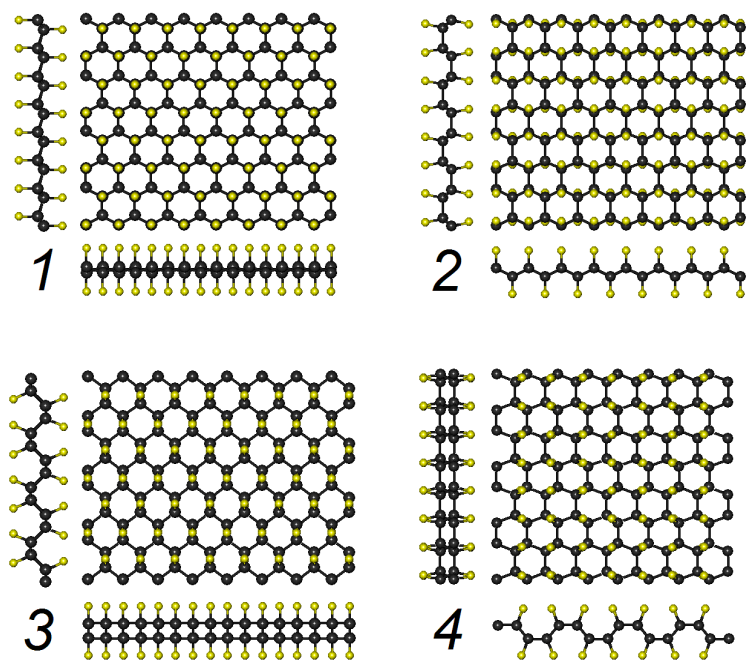

**Fig. 1.** Four different isomers of fluorographene CF: *1 – chair* with a chair conformation of $C_6$ ring, *2 – boat* with a boat conformation of $C_6$ ring, *3 – zigzag* with a chair conformation of $C_6$ ring, *4 – armchair* with a boat conformation of $C_6$ ring. C and F atoms are black and yellow, respectively.

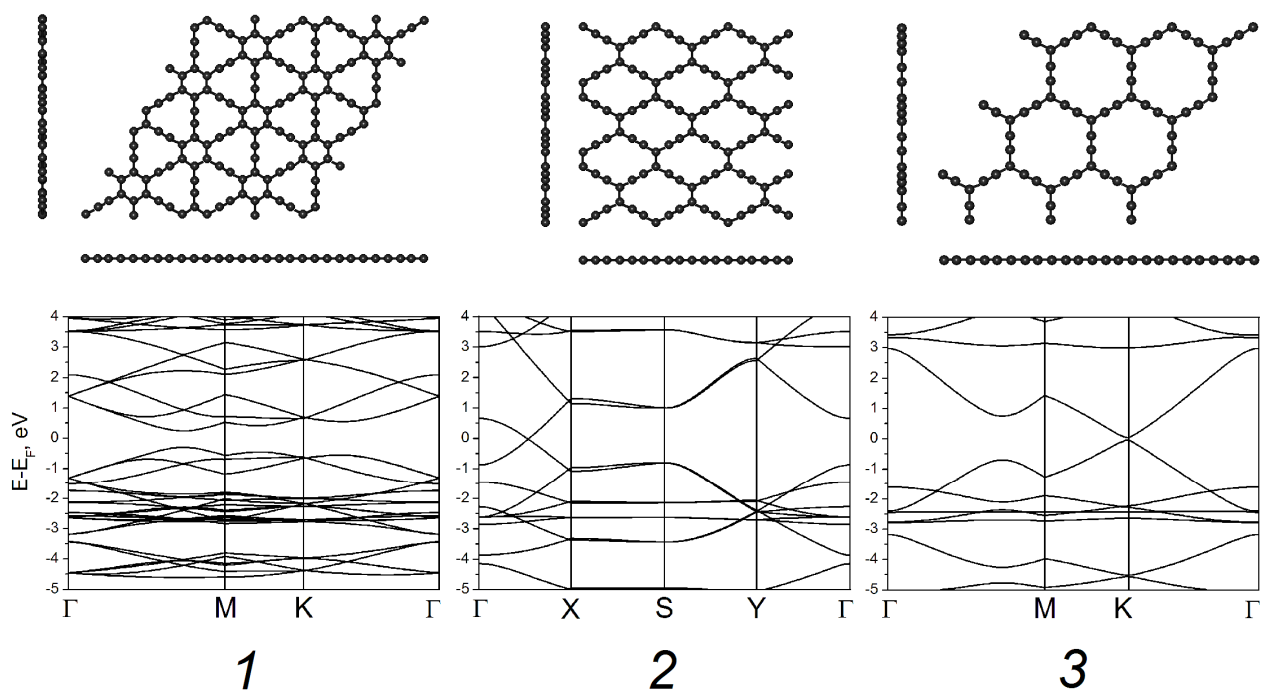

**Fig. 2.** Optimized atomic structures and electronic bands of the examined graphyne allotropes composed of: *1* - benzene rings connected by C-C dimers (GP*1*), *2* - ethylene units connected via C-C dimers (GP*2*), and *3* - carbon atoms connected by C-C dimers (GP*3*).

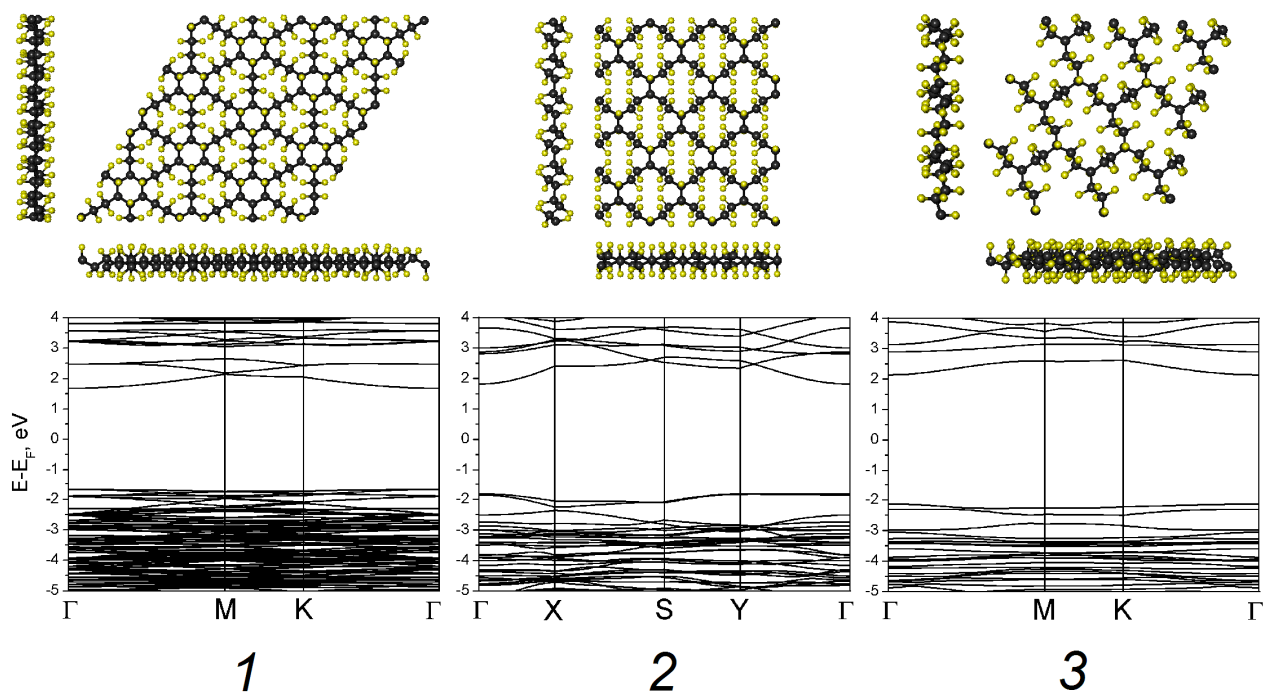

**Fig. 3.** Optimized atomic structures and electronic bands of the most stable isomers of fluorographynes *1*: FGP-$C_2F_3$; *2*: FGP-$C_3F_5$; and *3*: FGP-$C_4F_7$. C and F atoms are black and yellow, respectively.

# *Supporting Information*

The optimized structures of fluorinated graphynes GP*1* – GP*3* (see Fig. 2)

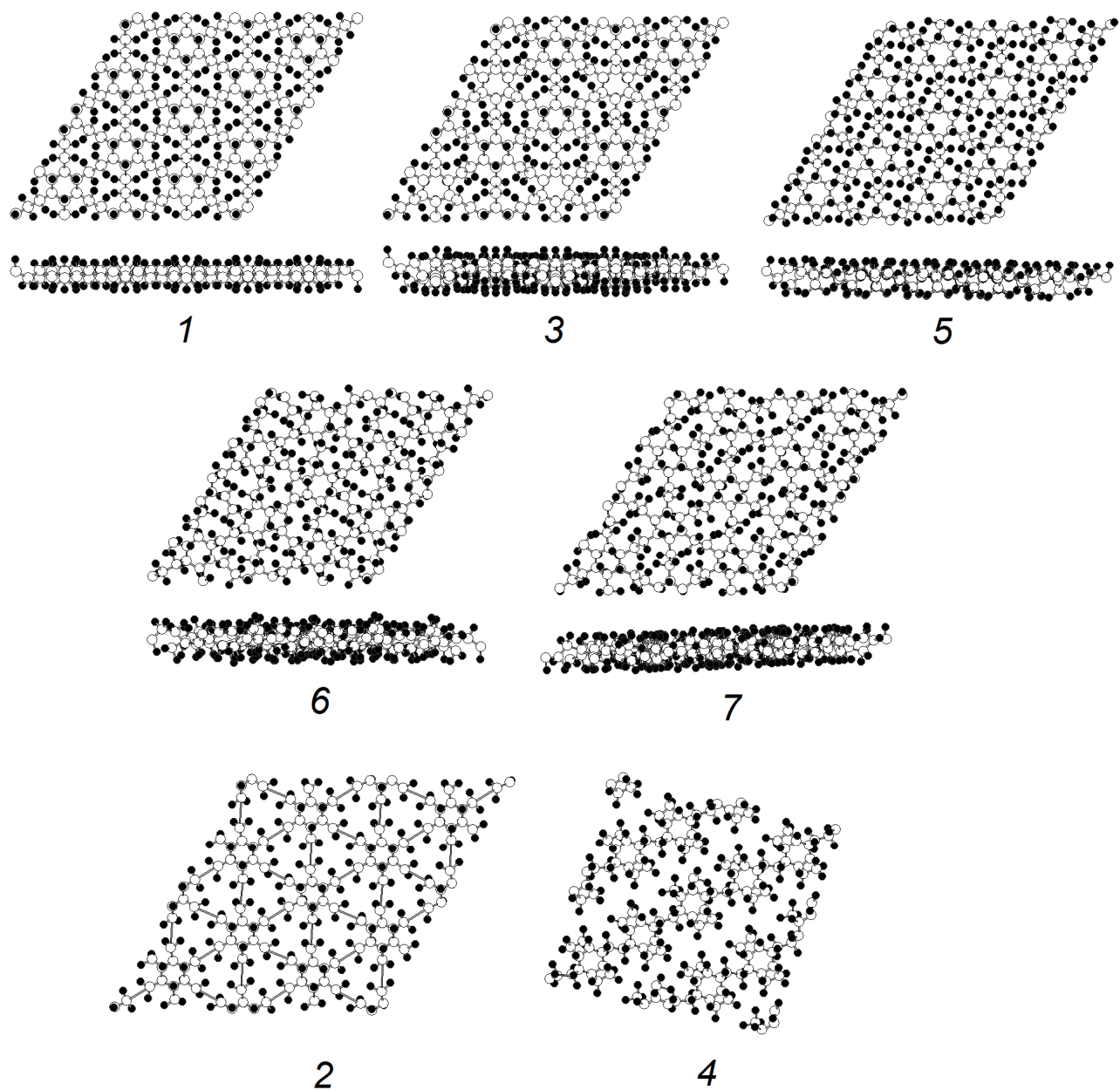

**Fig. S1.** Isomers and conformers of fluorinated graphyne GP*1* (FGP-$C_2F_3$) composed of benzene rings connected by C-C dimers. C and F atoms are painted in white and black, respectively. For unstable isomers *2* and *4* the initial structures are also given.

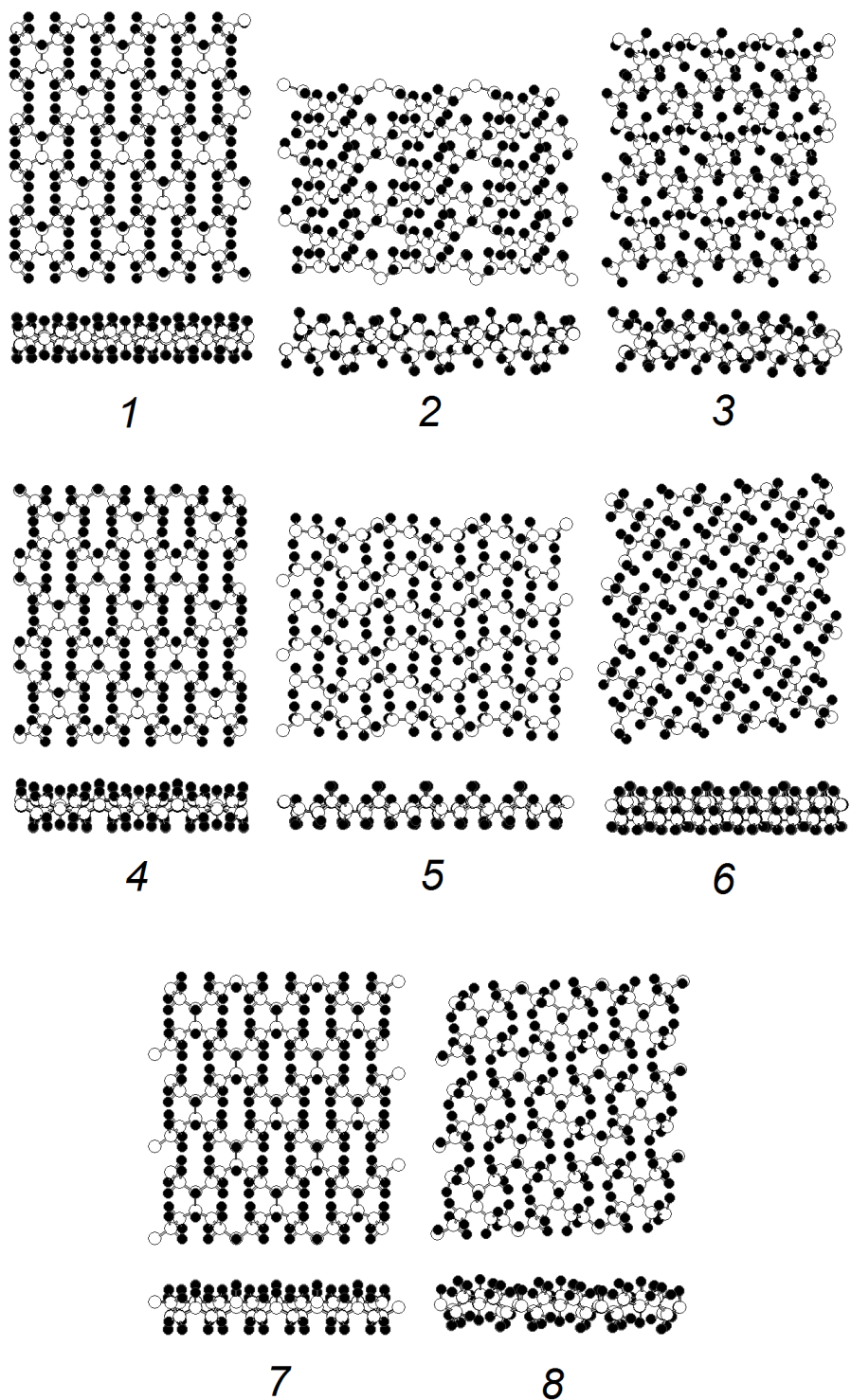

**Fig. S2.** Isomers and conformers of fluorinated graphyne GP*2* (FGP-C$_3$F$_5$) composed of ethylene units connected by C-C dimers. C and F atoms are painted in white and black, respectively.

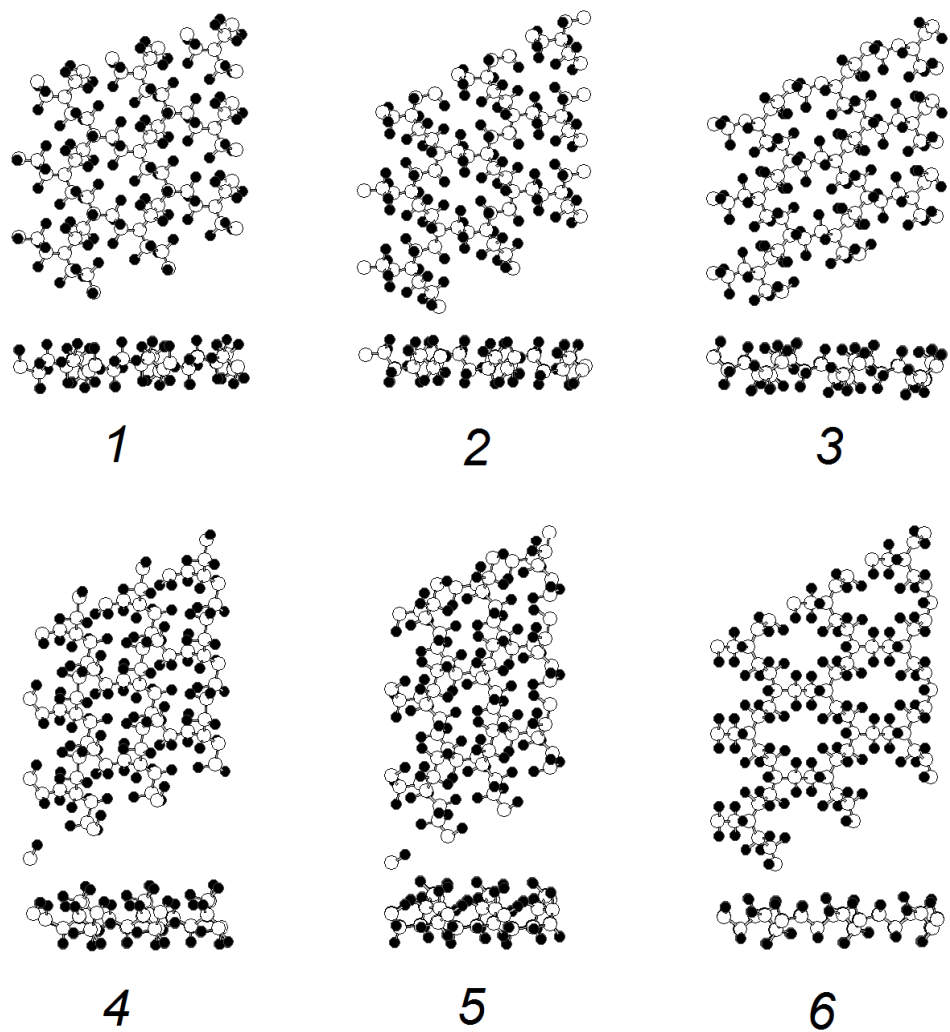

**Fig. S3.** Isomers and conformers of fluorinated graphyne GP*3* (FGP-C$_4$F$_7$) composed of carbon atoms connected by C-C dimers. C and F atoms are painted in white and black, respectively.